\documentclass[prx,twocolumn,groupedaddress]{revtex4}

\usepackage{graphicx}
\usepackage{bm}
\usepackage{braket}
\usepackage{physics}
\usepackage{bbold}
\usepackage{cases}
\usepackage{tabularx}
\usepackage{multirow}
\usepackage{braket}
\usepackage{epstopdf}
\usepackage[urlcolor=blue]{hyperref}

\newcommand{\be}{\begin{equation}}
\newcommand{\ee}{\end{equation}}

\newcommand{\bea}{\begin{eqnarray}}
\newcommand{\eea}{\end{eqnarray}}
\usepackage{color}
\usepackage{dcolumn}
\usepackage{epsfig}
\usepackage{bm}
\usepackage[urlcolor=blue]{hyperref}
\hypersetup{backref, colorlinks=true, linkcolor=blue, citecolor=blue}

\begin{document}

\title{Surface supercurrent diode effect}

\author{Noah F. Q. Yuan}
\email{fyuanaa@connect.ust.hk}
\affiliation{School of Science, Harbin Institute of Technology, Shenzhen 518055, China}

\begin{abstract}
We propose a new type of supercurrent diode effect on the surface of a superconductor with surface states under in-plane magnetic fields. Surface supercurrent diode effect can lead to a perfect supercurrent diode in a considerably wide range of fields. For comparison, the conventional supercurrent diode effect due to the spin-orbit coupling in a two-dimensional superconductor cannot be perfect in usual cases. 
Candidates such as the (001) surface of iron-based superconductors BaFe$_{2-x}$Co$_x$As$_2$ are discussed.
Calculations are performed under the Ginzburg-Landau formalism.
\end{abstract}

\maketitle

When both inversion and time-reversal symmetries are broken, the critical current of a superconductor can be nonreciprocal, known as the supercurrent diode effect (SDE). To theoretically elaborate the mechanism of SDE and experimentally demonstrate SDE turn out to be not any easy task. 

Experimentally, nonreicprocal transport in superconducting systems has been found in the fluctuating region \cite{Waka,Yasu,Qin,Hoshi}, where resistance is in general finite and nonlinear transport is significantly enhanced compared to the normal conducting state. Recently, Ando \textit{et. al.} \cite{Ando} realized a superconducting diode that has zero resistance in one direction but finite in the opposite direction, which inspired tremendous discoveries of SDEs in various superconducting systems, such as two-dimensional electron gas (2DEG) in quantum wells \cite{Chris}, transition metal dichalcogenides \cite{Banabir,Lorenz}, twisted bilayer graphene \cite{Diez} and twisted trilayer graphene \cite{LinJ}. 

As far as we know, SDE was first theoretically proposed by Victor M. Edelstein \cite{EdelEE,EdelEE1,EdelME,EdelSDE}. After the discovery of the famous Edelstein effect \cite{EdelEE,EdelEE1}, Edelstein proposed the magnetoelectric effect \cite{EdelME} and the supercurrent diode effect \cite{EdelSDE} for polar superconductors. As will be elaborated Section \ref{CSDE}, Edelstein's physical picture is correct while the detailed calculation was in fact inadequate to show SDE due to the gauge symmetry. 

Recently theories of SDE were developed \cite{Yuan,Akito,James,Harley,Zhai,Ilic}, including the one by Liang Fu and the author \cite{Yuan}. 
Such theories apply to two-dimensional (2D) superconductors and can be unified in the Ginzburg-Landau formalism, where Cooper pairs are boosted to finite momentum by external magnetic fields, so that critical currents parallel and anti-parallel to the Cooper pair momentum can be unequal. In order to capture the critical current nonreciprocity, the free energy is expanded to the quartic order in Cooper pair momentum $q$. In this work, we refer to this theory as the conventional theory of SDE, and this paper is devoted to address two issues related to it. 

\begin{figure}
\includegraphics[width=\columnwidth]{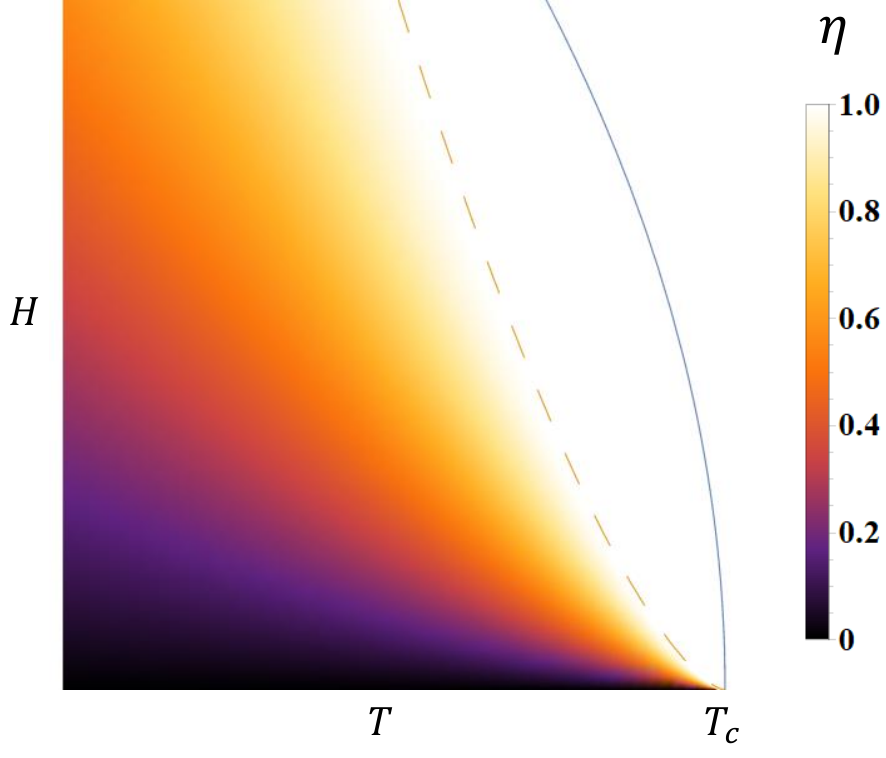}
\caption{Supercurrent diode coefficient $\eta$ under in-plane magnetic field $H$ and temperature $T$ in the surface supercurrent diode effect, when field and current are perpendicular. The solid blue line denotes the in-plane critical field $H_{c3\parallel}$ for surface superconductivity, and the dashed orange line denotes the field strength $H_{\rm M}$ beyond which $\eta=1$ (see maintext).}\label{fig1}
\end{figure}

First, the microscopic mechanism of the conventional SDE theory should be more elaborated. 
As elaborated in Sec. \ref{CSDE}, we show that the Edelstein effect is the microscopic driving force of the conventional supercurrent diode effect in two-dimensional superconductors with spin-orbit coupling (SOC).
Second, mechanisms other than the conventional theory should be explored for SDE. 
In Sec. \ref{SSDE}, we consider a three-dimensional (3D) superconductor with surface states, and find that the Meissner effect of surface states can also lead to SDE on its surface. 

In this work, we find the conventional supercurrent diode effect (CSDE) carried by conventional 2DEG with SOC and the surface supercurrent diode effect (SSDE) carried by surface states of a 3D superconductor.
In both types of SDE, the critical currents $J_{c}^{\pm}$ along opposite directions can be unequal, quantified by the dimensionless supercurrent diode coefficient
\bea
\eta\equiv\frac{J_{c}^{+}-J_{c}^{-}}{J_{c}^{+}+J_{c}^{-}}\in [-1,1].
\eea


In Fig. \ref{fig1} and Fig. \ref{fig2} the supercurrent diode coefficients of SSDE and CSDE are plotted in the field-temperature plane respectively, whose mechanisms and properties will be elaborated in the following two Sections.


\section{Conventional supercurrent diodes}\label{CSDE}
\subsection{Phenomenological theory}
The free energy of a 2D superconductor per area is
\begin{eqnarray}
f_{2{\rm D}}=\int \frac{d^2\bm q}{(2\pi)^2}\left\{\sigma|\Delta_{\bm q}|^2+\frac{1}{2}\gamma |\Delta_{\bm q}|^4\right\},
\end{eqnarray}
up to the quartic order of the order parameter $\Delta_{\bm q}$ with Cooper pair momentum $\bm q$, with quadratic and quaric coefficients $\sigma,\gamma$ respectively.
In order to describe SDE, the quartic expansion of $\sigma$ is carried out
\begin{eqnarray}\label{eq_alpha}
\sigma&=&a+b(\hat{\bm n}\times\bm H)\cdot\bm q,\\\nonumber
a&=&t+a_0 q^2-a_1 q^4,\\\nonumber
b&=&b_0 -b_1 q^2,
\end{eqnarray}
where $\bm H$ is the in-plane external magnetic field, $\hat{\bm n}$ is the normal vector of the 2D superconductor, $t=(T-T_c)/T_c$ is the reduced temperature with critical temperature $T_c$, and $a_{0,1},b_{0,1},\gamma$ are coefficients to be given later. The expansion of $\gamma$ in terms of $q$ may also be carried out \cite{Ilic}, which leads to higher order corrections.

By minimizing $f$, equilibrium Cooper pairs are found to have the finite momentum boosted by $\bm H$
\be
\bm q_0=\frac{b_0}{2a_0}\bm H\times\hat{\bm n}.
\ee

In the presence of a gauge potential $\bm A$, the free energy is changed upon minimal coupling $\bm q\to\bm q-2e\bm A$, and the supercurrent line density can be computed $\bm J=-\partial f_{2{\rm D}}/\partial\bm A$. In the 2D limit under in-plane fields we take the limit $\bm A\to\bm 0$, the supercurrent thus reads $\bm J=\int \frac{d^2\bm q}{(2\pi)^2}\bm J_{\bm q}$, where $\bm J_{\bm q}=e|\sigma|\partial_{\bm q}\sigma/\gamma $ is a function of $\bm q$, defined in the range such that $\sigma\leq 0$.

Although $q$-linear term is sufficient to describe finite-momentum superconductivity at zero current, higher order $q$-terms are needed for the critical current to be nonreciprocal.
When $\sigma=t+a_0q^2+b_0(\hat{\bm n}\times\bm H)\cdot\bm q$ is truncated at $q^2$-term, although $\sigma(\bm q)\neq \sigma(-\bm q)$ is not an even function of $\bm q$, it is found that $\sigma(\bm q_0+{\bm q})=\sigma(\bm q_0-{\bm q})$ with respect to the equilibrium momentum $\bm q_0$.
As a result, $\bm J_{\bm q_0+{\bm q}}=-\bm J_{\bm q_0-{\bm q}}$.
Such symmetry guarantees the critical current reciprocity $J_c^{+}=J_{c}^{-}$ and $\eta=0$.

To describe SDE, we consider the following expansion near the equilibrium momentum
\begin{eqnarray}
\sigma(\bm q+\bm q_0)=a-b'_1 q^2(\hat{\bm n}\times\bm H)\cdot\bm q,\quad
b'_1=b_1-2\frac{a_1}{a_0}b_0,
\end{eqnarray}
and find that the finite-momentum Cooper pairs boosted by external field would change the odd order coefficients $b_0\to 0,b_1\to b'_1$.
Due to this modification, up to the linear order in $\bm H$, the supercurrent diode coefficient reads
\be\label{eq_eta0}
\eta=\left(\frac{b_1}{b_0}-2\frac{a_1}{a_0}\right)\sqrt{\frac{|t|}{3a_0^3}}b_0{(\bm H\times\hat{\bm n})\cdot\hat{\bm i}},
\ee
where $\hat{\bm i}$ is the current direction.
Details of the calculations can be found in the Appendix.

\begin{figure}
\includegraphics[width=\columnwidth]{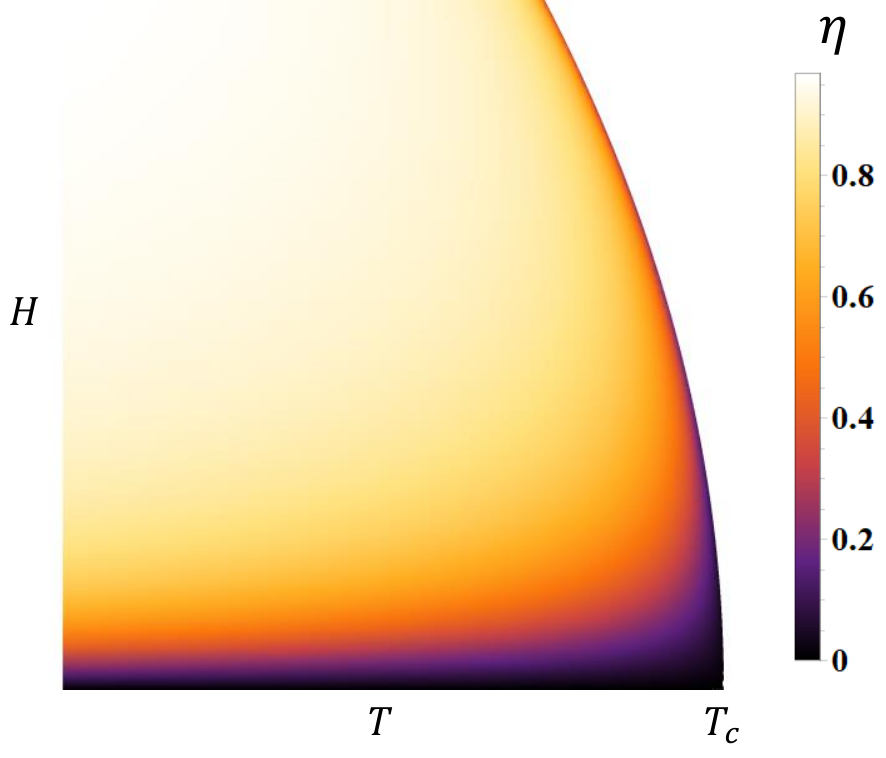}
\caption{Supercurrent diode coefficient $\eta$ under in-plane magnetic field $B$ and temperature $T$ in the conventional supercurrent diode effect of a Rashba superconductor, when the supercurrent and the magnetic field are perpendicular.}\label{fig2}
\end{figure}

\subsection{Microscopic theory}
Next we compute the Ginzburg-Landau coefficients $a_{0,1},b_{0,1},\gamma$ in Eq. (\ref{eq_alpha}) from a microscopic model of the Rashba superconductor \cite{Yuan}.
Under an in-plane magnetic field $\bm H$, the orbital effect is screened, and the Zeeman effect dominates.
Denote $c_{s}$ as the electron with spin $s=\uparrow,\downarrow$, then on the Nambu basis $(c_{\uparrow},c_{\downarrow},-c^{\dagger}_{\downarrow},c^{\dagger}_{\uparrow})$, the Bogouliubov-de Gennes (BdG) Hamiltonian reads
\bea
\mathcal{H}&=&
\begin{pmatrix}
H_{\bm k+\frac{1}{2}\bm q} & \Delta_{\bm q} \\
\Delta_{\bm q}^{*} & -\mathcal{T} H_{-\bm k+\frac{1}{2}\bm q}\mathcal{T}
\end{pmatrix},\\
H_{\bm k}&=&\frac{|\bm k|^2}{2m}-\varepsilon_{\rm F} + \alpha_{\rm R}(\bm k\times\bm\sigma)\cdot\hat{\bm n}+\mu_{\rm m}\bm H\cdot\bm\sigma,
\eea
where $H_{\bm k}$ describes the normal state electrons with momentum $\bm k$, spin $\bm\sigma$, mass $m$, Fermi energy $\varepsilon_{\rm F}$, Rashba spin-orbit coupling (SOC) strength $\alpha_{\rm R}$ and the magnetic moment $\mu_{\rm m}$. Electrons form Cooper pairs with momentum $\bm q$ under the pairing potential $\Delta_{\bm q}$, and $\mathcal{T}=i\sigma_y K$ is the antiunitary time-reversal operator with complex conjugation $K$. 

To the linear order in $\bm q$, we have
\be
\mathcal{H}=H_{0}\tau_z +\Delta_{\bm q}\tau_x+\mu_{\rm m}\bm H\cdot\bm\sigma +\frac{\bm k\cdot\bm q}{2m}
+\frac{\alpha_{\rm R}}{2}(\bm q\times\bm\sigma)\cdot\hat{\bm n}.
\ee
The zero-field normal Hamiltonian $H_0=H_{\bm k}|_{\bm H=\bm 0}$
gives rise to two energy bands $\xi_{\pm}={|\bm k|^2}/({2m})-\varepsilon_{\rm F}\pm \alpha_{\rm R} |\bm k|$ split by SOC. Correspondingly the two Fermi surfaces $\xi_{\pm}=0$ will have the same Fermi velocity $v_{\rm F}=\sqrt{2\varepsilon_{\rm F}/m+\alpha_{\rm R}^2}$ but different density of states (DOS) $N_{\pm}=N_0(1\mp {\alpha_{\rm R}}/{v_{\rm F}})$ with $N_0=m/(2\pi)$ \cite{EdelME,Samokhin,Dimi1,Dimi2}.

There are three depairing terms, from the Zeeman effect $\mu_{\rm m}\bm H\cdot\bm\sigma$, the Doppler effect $ \bm k\cdot\bm q/(2m) $ of the kinetic energy, and most importantly the Edelstein effect $\frac{1}{2}{\alpha_{\rm R}}(\bm q\times\bm\sigma)\cdot\hat{\bm n}$ of the SOC \cite{EdelEE,EdelEE1}, respectively. 


The even order coefficients $a_{0,1}$ and the coefficient $\gamma$ can be derived as in the conventional BCS theory
\begin{eqnarray}
a_0=\frac{1}{4}\frac{C_0v_{\rm F}^2}{(\pi T_c)^2},\; a_1=\frac{1}{8}\frac{C_1 v_{\rm F}^4}{(\pi T_c)^4},\; \gamma=\frac{C_0}{(\pi T_c)^2},
\end{eqnarray}
where $C_0=7\zeta(3)/8$, $C_1=93\zeta(5)/2^8$ are numerical constants, $v_{\rm F}=\sqrt{2\varepsilon_{\rm F}/m+\alpha_{\rm R}^2}$ is the Fermi velocity.
The odd order coefficients $b_{0,1}$ are determined by SOC, through both band splitting effect and Edelstein effect:
\begin{eqnarray}\label{eq_b}
b_0=\frac{3}{2}\frac{C_0\alpha_{\rm R}}{(\pi T_c)^2}\mu_{\rm m},\; b_1=\frac{5}{2}\frac{C_1 v_{\rm F}^2\alpha_{\rm R}}{(\pi T_c)^4}\mu_{\rm m}.
\end{eqnarray}
From the coefficients above, we can work out
the conventional supercurrent diode coefficient
\be\label{eq_etac}
\eta ={0.64}\frac{\alpha_{\rm R}}{v_{\rm F}}\frac{(\bm H\times\hat{\bm n})\cdot\hat{\bm i}}{H_P}\sqrt{1-\frac{T}{T_c}},
\ee
where $H_P=1.25T_c/\mu_{\rm m}$ is the Pauli limit. 
This formula has been numerically verified in Ref. \cite{Yuan} with the correct numerical factor.

There have been a series of works in deriving the correct formula for the SDE. As far as we know, Victor M. Edelstein himself was one of the pioneers to theoretically study the possible SDE in a superconductor without inversion symmetry under an external magnetic field \cite{EdelSDE}. 
However, the expansion of $\sigma$ was truncated at $q^2$ term, and hence was not able to describe the CSDE as elaborated above.
Nevertheless, in Ref. \cite{EdelSDE}, the Ginzburg-Landau coefficient $b_0$ is the same as that in Eq. (\ref{eq_b}).


Liang Fu and the author studied the SDE of a Rashba superconductor (which we dub as CSDE) in Ref. \cite{Yuan} by both numerical and analytical methods. The numerical model includes the full effect of Zeeman field, the Doppler effect and the Edelstein effect, hence numerical calculations lead to the supercurrent diode coefficient consistent with Eq. (\ref{eq_etac}). However, in the analytical calculations, the Edelstein effect was absent and the the expansion of $\sigma$ was truncated at $q^3$ term only. 
Interestingly, the analytically derived supercurrent diode coefficient was still the same as Eq. (\ref{eq_etac}), in spite of the absence of two ingredients in the expansion of free energy ($q^4$ term and Edelstein term).

Recently, S. Ili\'c and F. S. Bergeret further studied the SDE \cite{Ilic} by expand $\sigma$ up to $q^4$ term. 
It was then found that the supercurrent diode coefficient was zero to the linear order of the field without the Edelstein effect.

The Edelstein effect is in fact the driving force of the CSDE. 
Due to the Edelstein effect, Cooper pairs with momentum $\bm q$ will also carry spin polarization $\bm S\propto\hat{\bm n}\times\bm q$ \cite{EdelME}. The external magnetic field $\bm H$ then couples to Cooper pairs via the Zeeman effect $\bm S\cdot\bm H\propto(\hat{\bm n}\times\bm H)\cdot\bm q$ and hence favors a particular direction for $\bm q$ and for the supercurrent, leading to the CSDE.
To explicitly demonstrate this, one can calculate the odd order coefficients without the Edelstein effect, which leads to smaller values
$b_0=C_0{\alpha_{\rm R}}\mu_{\rm m}/{(\pi T_c)^2},\; b_1= C_1{v_{\rm F}^2\alpha_{\rm R}}\mu_{\rm m}/{(\pi T_c)^4}$ compared with those in Eq. (\ref{eq_b}). 
Since $b_1/b_0=2a_1/a_0$ without the Edelstein effect,
the supercurrent diode coefficient vanishes $\eta\equiv 0$ according to Eq. (\ref{eq_eta0}).
Moreover, the critical field diverges at zero temperature when the Edelstein effect is not included \cite{Samokhin,Dimi1,Dimi2}. 
To have nonzero supercurrent diode effect at finite temperatures and finite critical field at zero temperature, it would be necessary to include the Edelstein effect in a consistent microscopic theory of the two-dimensional Rashba superconductors.


\section{Surface supercurrent diodes}\label{SSDE}
\subsection{Ginzburg-Landau formalism}
To improve the supercurrent diode coefficient, we can consider strong-SOC materials. In particular, we can consider the strongest limit of SOC $\alpha_{\rm R}\to v_{\rm F}$, which corresponds to the surface states described by a single Dirac cone Hamiltonian $H_{\bm k}=v_{\rm F}(\bm k\times\bm\sigma)_z$, so that $\alpha_{\rm R}=v_{\rm F}$. 

However, a single Dirac cone cannot exist by itself, but can live with the whole bunch of bulk states. As a result, to describe the supercurrent diode effect of superconducting surface states, one may include the bulk superconductor as well.

We consider a semi-infinite superconductor in the lower half space $z\leq 0$, which its normal conducting phase possesses surface states. 
In the Ginzburg-Landau (GL) regime, the temperature and the field are near the superconducting phase transitions and the length scales are greater than the zero-temperature bulk coherence length. Since surface states usually extend to the bulk by a few atomic layers \cite{TI}, in the GL regime the surface states only live on the surface $z=0$ of the superconductor. After washing out short-range details, the GL free energy is composed of the bulk term and the surface term
\be
    F=\int_{V} f_{3{\rm D}}d^3\bm r+\int_{\partial V}f_{2{\rm D}} d^2\bm r,
\ee
where the bulk energy density is
\be
    f_{3{\rm D}}=\alpha|\psi|^2+\frac{\beta}{2}|\psi|^4+\frac{|\bm D\psi|^2}{2m}+\frac{1}{2}(\bm B-\bm H)^2,
\ee
and the surface energy density is
\be
    f_{2{\rm D}}=\sigma |\psi|^2+\frac{1}{2}\gamma |\psi|^4.
\ee
Here, $\psi(\bm r)$ is the superconducting order parameter, $\bm D=\nabla-2ie\bm A$ is the covariant derivative, and $\bm A(\bm r)$ is the vector potential. 
Due to the Meissner effect in 3D superconductors, besides the internal field $\bm B=\nabla\times\bm A$, we also introduce the external field $\bm H$.

The bulk free energy density $f_{3{\rm D}}$ is characterized by three parameters $m,\alpha$ and $\beta$, where $m$ is the electron mass, $\alpha=\alpha_0(T-T_{c0})$ is a function of temperature $T$, and $T_{c0}$ is the bulk critical temperature. The stability of the bulk free energy requires $m,\alpha_0,\beta$ to be all positive.

The surface free energy density $f_{2{\rm D}}$ is characterized by the quadratic surface energy $\sigma$ \cite{Noah,Simonin,Andryushin,deGennes,saint} and quartic term $\gamma$ as described in Section \ref{CSDE}. 
As the bulk free energy describes the second-order phase transition between two phases (i.e. superconducting versus normal), in the boundary free energy the quartic term is irrelevant \cite{Diehl} and hence can be neglected in the rest of this manuscript. 

When surface states are present, the surface energy is negative $\sigma<0$, suggesting the preferential accumulation of Cooper pairs on the surface. 
As a result, the superconducting phase transition consists of two steps. The first step is at temperatures above the bulk critical temperature, known as the superconducting nucleation, where the electrons condense into Cooper pairs in the surface layer, while the bulk electrons stay mostly normal. In the second step, the temperature is lower than the bulk critical temperature, the bulk is superconducting as well as the surface layer. 

By minimizing the GL free energy with respect to $\psi$, we derive the GL equation
\bea\label{eq_GLs}
\left(-\frac{\bm D^2}{2m}+\alpha +\beta|\psi|^2\right)\psi =0,\ (\bm r\in V)
\eea
and the de Gennes boundary condition \cite{Noah,Simonin,Andryushin,deGennes,saint}
\bea\label{eq_dG}
    \hat{\bm n}\cdot\bm D\psi =\frac{\psi}{l_s},\ (\bm r\in\partial V)
\eea
where $l_{s}=1/(2m|\sigma|)>0$ is the extrapolation length denoting the thickness of surface superconductivity, and $\hat{\bm n}$ is the normal vector of the surface pointing outward.

By minimizing the GL free energy with respect to $\bm A$, the Ampere's law is derived with supercurrent density $\bm j$
\bea\label{eq_GLv}
\nabla\times\bm B=\bm j\equiv\frac{2e}{m} {\rm Im}(\psi^{*}\bm D\psi)\ (\bm r\in V).
\eea
According to the de Gennes boundary condition in Eq. (\ref{eq_dG}), no bulk supercurrent penetrates through the surface $\hat{\bm n}\cdot\bm j|_{\partial V}=0$, since $\psi^{*}\hat{\bm n}\cdot\bm D\psi$ is real on the surface.


\subsection{Surface superconductivity}
Near the superconducting phase transition, $|\psi|$ will be small and the GL equation Eq. (\ref{eq_GLs}) can be linearized. We then find the solutions to the linearized GL equation to obtain properties on the superconducting phase transitions. Details on the critical temperature and critical field at zero current can be found in Ref. \cite{Noah}. We review the main results below.

At zero field and zero current, the surface superconductivity is confined within the surface layer of thickness $\sim l_s$, and the onset critical temperature for the surface superconductivity is $T_c =T_{c0} +({2m\alpha_0 l_s^2})^{-1}$, higher than the bulk critical temperature. Near $T_c$, we find
\be\label{eq_phi}
\psi=\phi\exp\left(-\frac{|z|}{l_s}\right),
\ee
where $\phi=\sqrt{{2\alpha_0(T_c-T)}/{\beta}}$ is determined by the nonlinear GL equation as shown in the Appendix.

Under finite fields and zero current, superconductivity is in general suppressed by the fields. It is well known that the upper critical field for the bulk is $H_{c2}=\Phi_0/(2\pi\xi_{0}^2)\propto (T_{c0}-T)^1$, where $\xi_{0}(T)\equiv\sqrt{1/(2m|\alpha|)}$ is the bulk coherence length and $\Phi_0=h/(2e)$ is the flux quantum \cite{Abrikosov1}. When $H>H_{c2}$, bulk superconductivity is killed, while surface superconductivity can still survive until a new critical field, which is the critical field $H_{c3}$ of the surface superconductivity. For out-of-plane fields $H_{c3\perp}=\Phi_0/(2\pi\xi^2)\propto (T_c-T)^1$, where the coherence length is $\xi\equiv 1/\sqrt{2m\alpha_0(T_c-T)}$, while for in-plane fields $H_{c3\parallel}=0.525{\Phi_0}/{\left(\xi^2 l_s\right)^{\frac{2}{3}}}\propto (T_c-T)^{2/3}$ \cite{Noah}.

In this work, we consider weak in-plane magnetic fields. In conventional superconductors we would expect Meissner effect, where magnetic fields are repelled from the bulk. As a result, the magnetic field can only penetrate into the surface layer, and supercurrent can only flow on the surface layer as well. The thickness of this surface layer is another fundamental length scale of the superconductor known as the London penetration depth $\lambda_L\equiv (2e)^{-1}\sqrt{m\beta/|\alpha|}\propto|T_{c0}-T|^{-1/2}$.

However, in our case of superconductors with surface states, when $T_{c0}<T<T_c$ superconductivity only survives in the surface layer of thickness $\sim l_s$ already, and the Meissner effect is different. To see this, we plug in the order parameter Eq. (\ref{eq_phi}) of the surface superconductivity into the Ampere's law Eq. (\ref{eq_GLv}), then near $T_c$ 
\be
\bm B=\bm H \exp\left(-\frac{2|z|}{l_s}\right).
\ee
It can be seen that the London penetration depth $\lambda_{L}$ does not appear in the spatial distribution of the internal magnetic field $\bm B$ at least to the leading order. Instead, the magnetic field fully penetrates into the surface layer of thickness $\sim l_s$ where surface superconductivity lives. Deep in the bulk, both magnetic field and superconductivity vanish. This is the Meissner effect in the case of superconductors with surface states.

To be concrete, one can calculate that at zero external current but finite field $\bm H$, surface Cooper pairs have zero momentum, while a persistent current $\bm J_0=\int_{-\infty}^{0}\bm j dz=\hat{\bm n}\times\bm H$ will be induced to screen magnetic field in the bulk. 
Notice that the supercurrent line density $\bm J_0$ has the same dimension as the magnetic field $\bm H$.
At finite external current and field, surface Cooper pairs can have finite momentum $\bm q$, and the supercurrent line density is
\bea\label{eq_Kt}
\bm{J}=\int_{-\infty}^{0}\bm j dz =\left(I_0\bm q+\bm J_0\right)(1-\xi^2 q^2),
\eea
where $\bm q$ is confined in the range $q<1/\xi$ such that the system remains superconducting, and
\be\label{eq_qt}
I_0=\frac{l_s}{\kappa^2}H_{c3\perp},\quad \bm J_0=\hat{\bm n}\times\bm H.
\ee
Notice that $H_{c3\perp}=\Phi_0/(2\pi\xi^2)$ is the out-of-plane critical field of the surface superconductivity with the coherence length $\xi\equiv 1/\sqrt{2m\alpha_0(T_c-T)}$ and the bulk GL parameter $\kappa=\lambda_{L}/\xi_{0}$, and $\bm J_0$ is the field-induced current at $\bm q=\bm 0$.


We can then calculate the critical currents parallel and anti-parallel to a given direction $\hat{\bm i}$ by maximizing and minimizing $\bm J\cdot\hat{\bm i}$ over $\bm q$ in the range $q<1/\xi$ respectively. Namely, $J_{c}^{+}={\rm max}(\bm J\cdot\hat{\bm i})$ and $J_{c}^{-}=-{\rm min}(\bm J\cdot\hat{\bm i})$. At weak field, we find $J_{c}^{\pm}=J_c(1\pm\eta)$, the average is $J_c=\frac{2\sqrt{3}}{9}I_0/\xi$, and the surface supercurrent diode coefficient reads
\bea\label{eq_etas}
\eta=\sqrt{3}\kappa^2\frac{(\bm H\times\hat{\bm n})\cdot\hat{\bm i}}{H_{c3\perp}}\frac{\xi}{l_s}.
\eea
This surface supercurrent diode coefficient $\eta$ is plotted in the phase plane of magnetic field $H$ and temperature $T$, with supercurrent perpendicular to the field, as shown in Fig. \ref{fig1}. At weak fields, $\eta\propto H(T_c-T)^{-3/2}$ and $J_c\propto H(T_c-T)^{3/2}$, hence the critical current difference $J_{c}^{+}-J_{c}^{-}\propto H$ is temperature independent.
At higher fields when $H\geq H_{\rm M}\equiv(l_s/\xi)H_{c3\perp}/\kappa^2$ we find $\eta=1$ until the critical field $H_{c3\parallel}$ where surface superconductivity vanishes. A perfect supercurrent diode hence could be realized in the field regime $H_{\rm M}<H<H_{c3\parallel}$ on the surface of a superconductor with surface states, as shown in Fig. \ref{fig1} where the Meissner field $H_{\rm M}\propto(T_c-T)^{3/2}$ is denoted by the dashed orange line and the critical field of surface superconductivity $H_{c3\parallel}\propto(T_c-T)^{2/3}$ is denoted by the solid blue line.

For further lower temperatures $T<T_{c0}$, the surface together with the bulk are all superconducting. Since the surface states contribution now becomes negligible compared with the bulk, the supercurrent becomes reciprocal deep in the bulk superconducting state.
Without surface states, the surface energy is zero $\sigma\to 0$, $l_s\to\infty$, and the critical current also becomes reciprocal according to Eq. (\ref{eq_etas}) \cite{Abrikosov2}. 
Thus to realize SSDE, surface states are needed on the surface of a superconductor and the temperature range is close to the onset critical temperature so that only surface superconductivity exists.

\subsection{Candidates}
We have considered an ideal model of the semi-infinite superconductor with surface states on its only one surface (the top surface). However, realistic materials will at least have two surfaces, the top and the bottom surfaces.
We suppose surface states exist on both top and bottom surfaces, with extrapolation lengths ${l_s^{\rm top}}$ and ${l_s^{\rm bot}}$ respectively, and states on the top surface are decoupled from the bottom surface.  
Without loss of generality, we assume $l_s^{\rm top}<l_s^{\rm bot}$, then the onset critical temperature for the surface superconductivity is $T_c =T_{c0} +({2m\alpha_0 |l_s^{\rm top}|^2})^{-1}$.
In the temperature regime $T_{c1}<T<T_{c}$ with $T_{c1}=T_{c0} +({2m\alpha_0 |l_s^{\rm bot}|^2})^{-1}$, only the top surface is superconducting while the bulk and the bottom surface are not. Eq. (\ref{eq_etas}) then applies in this temperature regime.
For lower temperatures $T_{c0}<T<T_{c1}$, both the top and the bottom surfaces are superconducting while the bulk is not, and the total surface supercurrent diode coefficient from these two surfaces is $\eta=\eta(l_s^{\rm top})-\eta(l_s^{\rm bot})$.

The iron-based superconductors FeSe$_{1-x}$Te$_{x}$ ($x\sim 0.5$) \cite{Gang,Zhang,Wang,Kong} provide candidates of superconductors with surface states and hence of SSDE. 
It has been theoretically calculated that \cite{Gang}, 
the $3d$ electrons from Fe atoms will exhibit band inversion in the bulk spectrum, 
and topological surface states would emerge on the (001) surfaces (top and bottom) as observed experimentally \cite{Zhang,Wang,Kong}. However, due to the topological protection, both the top and bottom surfaces will host surface states, and one cannot eliminate states on one surface while remaining another by local operations. Therefore, in usual samples of FeSe$_{1-x}$Te$_{x}$, we expect $l_s^{\rm top}\approx l_s^{\rm bot}$, and hence the SSDE would be weak, unless the top and bottom surfaces can be made sharply different.

Notice that our theory of SSDE in fact does not specify the type and origin of the surface states.
Surface states in SSDE may have various origins, such as topological surface states in FeSe$_{1-x}$Te$_{x}$ \cite{Gang,Zhang,Wang,Kong} as discussed above, surface reconstructions in cleaved or covered superconductors \cite{Gao,Heumen,Par} as will be discussed below, and other crystal defect states \cite{Matsu,Khl}.

We thus turn to superconductors with surface reconstruction, which fall into another family of iron-based superconductors, the doped iron arsenides BaFe$_{2-x}$Co$_x$As$_2$ \cite{Gao,Heumen,Par}. The surface Ba atoms predominantly favor a $\sqrt{2}\times\sqrt{2}$ order \cite{Gao}, which leads to the lattice reconstruction of the surface layer. As a result, surface related Fe 3$d$ states are present in the electronic structure in the form of surface bands as both theoretically calculated and experimentally observed \cite{Heumen}. By changing the surface condition upon cleavage or coverage, one can change the reconstruction and hence the onset critical temperature as observed experimentally \cite{Par}. Compared with topological surface states in FeSe$_{1-x}$Te$_{x}$, the surface reconstruction in BaFe$_{2-x}$Co$_x$As$_2$ can be well adjusted on just one surface. Consequently, the SSDE in BaFe$_{2-x}$Co$_x$As$_2$ is expected to be significant upon proper surface conditions.

Finally we address the dependence of supercurrent diode coefficient $\eta$ on field $\bm H$ and current $\hat{\bm i}$ in the two types of SDEs in this work, which turns out to be dictated by the underlying point group of the superconducting system.
Notice that $\eta$ is time-reversal even while $\bm H$ and $\hat{\bm i}$ are time-reversal odd, thus we expect $\eta$ to be expressed as a quadratic form of $\bm H$ and $\hat{\bm i}$, which is invariant under the point group.
On the surface of a three-dimensional superconductor or in the plane of a two-dimensional Rashba superconductor, the point groups are of the same type, namely $C_{nv}$ or its subgroup, where $n=2,3,4,6,\infty$. When $n>2$, $\bm H$ and $\hat{\bm i}$ furnish the same 2D irreducible representation, while the normal vector $\hat{\bm n}$ furnishes the 1D irreducible representation $A_2$. In order to furnish the trivial representation we find $\eta\propto ({\bm H}\times\hat{\bm n})\cdot\hat{\bm i}$ as shown in Eqs. (\ref{eq_etac}) and (\ref{eq_etas}).
In a two-dimensional superconductor without the inversion center, however, other quadratic invariants will become possible, since the point group is not limited to $C_{nv}$ \cite{Yuan,Yuan1}. For example $\eta\propto\bm H\cdot\hat{\bm i}$ when the point group is $D_{n}$ ($n=2,3,4,6$), and $\eta\propto\bm H\cdot\hat{\bm i}-2(\bm H\cdot\hat{\bm m})(\hat{\bm m}\cdot\hat{\bm i})$ when the point group is $D_{2d}$, where $\hat{\bm m}$ is the normal vector of the mirror plane.

\section{Conclusions}
In this work, we first show that in a two-dimensional superconductor with spin-orbit coupling, the Edelstein effect makes supercurrent spin-polarized and hence an external Zeeman field will favor the supercurrent along particular directions, leading to the conventional supercurrent diode effect.
Second, we find that in three-dimensional superconductors with surface states, close to the superconducting-normal phase transitions, superconductivity can only survive in the surface layer. An external in-plane field can generate a persistent current in the surface layer and hence affect the critical current of the surface superconductivity, resulting in the surface supercurrent diode effect. 
In the phase plane spanned by the temperature and the field, the supercurrent diode coefficient of CSDE decreases when approaching the critical field line (Fig. \ref{fig2}), while that of SSDE increases close to the superconducting-normal phase transition (Fig. \ref{fig1}).
Moreover, a perfect supercurrent diode can be realized in SSDE near the critical field regime, where currents along one direction can be supercurrent when smaller than the critical value, while along the opposite direction no current can become superconducting.
Candidates for SSDE are discussed.

\textit{Acknowledgement}---This work is supported by the National Natural Science Foundation of China (Grant. No. 12174021).


\appendix
\section{Surface supercurrent diode effect}
\subsection{Mean-field derivations of Ginzburg-Landau free enrgy} 
We consider the model Hamiltonian at zero field with effective on-site attractive interaction among electrons
\be
    H=\int d^3\bm r d^3\bm s c^{\dagger}(\bm r)\mathcal{H}(\bm r,\bm s)c(\bm s)
    -\frac{g}{2}\int d^3\bm r \left\{c^{\dagger}(\bm r)c(\bm r)\right\}^2,
\ee
where $c=\{c_{\mu}\}^{\rm T}$, $c_{\mu}(\bm r)$ denotes an electron at site $\bm r$ with band index $\mu$, $\mathcal{H}(\bm r,\bm s)$ is the normal state Hamiltonian matrix of the topological material, and $g>0$ is the on-site attraction strength between electrons.
We would like to compute the $s$-wave pairing correlation $\Delta(\bm r)=g\langle c_{\uparrow}(\bm r)c_{\downarrow}(\bm r)\rangle$ in order to obtain the physics of superconductivity in this system, where $\langle O\rangle\equiv {\rm Tr}(O e^{-\frac{H}{k_{\rm B}T}})/Z$ denotes the thermodynamic average at temperature $T$, and $Z\equiv {\rm Tr}e^{-\frac{H}{k_{\rm B}T}}$ is the partition function.
To capture the long-range physics of $\Delta(\bm r)$, we expand the free energy $F\equiv {-{k_{\rm B}T}}\log Z$ within mean field
\be
    F=\int d^3\bm r\frac{|\Delta(\bm r)|^2}{g}-\int d^3\bm r d^3\bm s K(\bm r,\bm s)\Delta^*(\bm r)\Delta(\bm s),
\ee
where $K$ is the kernel in terms of sum over Mastubara frequency $\omega=(2n+1)\pi k_{\rm B}T$ $(n\in\mathbb{Z})$:
\be
    K(\bm r,\bm s)=k_{\rm B}T\sum_{\omega ab}
    \frac{\phi_{a}(\bm r)^{\dagger}\phi_{a}(\bm s)}{\xi_a-i\omega} 
    \left[\frac{\phi_{b}(\bm r)^{\dagger}\phi_{b}(\bm s)}{\xi_b-i\omega}\right]^{*},
\ee
and eigenstates $\mathcal{H}\phi_a=\xi_a\phi_a$ of the normal Hamiltonian.
Deep in the bulk $|\bm r|\to\infty$, the kernel only involves bulk states $\phi^{\rm B}$ and hence has full translation symmetry
\bea\nonumber
    K^{\rm B}(\bm r,\bm s)&=&k_{\rm B}T\sum_{\omega ab}
    \frac{\phi^{\rm B}_{a}(\bm r)^{\dagger}\phi^{\rm B}_{a}(\bm s)}{\xi_a-i\omega} 
    \left[\frac{\phi^{\rm B}_{b}(\bm r)^{\dagger}\phi^{\rm B}_{b}(\bm s)}{\xi_b-i\omega}\right]^{*}\\
    &=&K^{\rm B}(\bm r-\bm s).
\eea
As a result, the free energy is composed of the bulk term and the surface term $F=F_0+F_s$
\bea\nonumber
    F_0&=&\int d^3\bm r\frac{|\Delta(\bm r)|^2}{g}-\int d^3\bm r d^3\bm s K^{\rm B}(\bm r-\bm s)\Delta^*(\bm r)\Delta(\bm s),\\\nonumber
    F_s&=&\int d^3\bm r d^3\bm s\Delta^*(\bm r)G(\bm r),\\\nonumber
    G&=&\int d^3\bm s [K^{\rm B}(\bm r-\bm s)-K(\bm r,\bm s)]\Delta(\bm s).
\eea
To rewrite the free energy within GL formalism, we introduce the order parameter
\be
\psi(\bm r)={\sqrt{mL}}\Delta(\bm r),\quad
L=\frac{1}{3}\int K^{\rm B}(\bm r)|\bm r|^2 d^3\bm r,
\ee
then the bulk free energy becomes the conventional GL free energy
\bea
    F_0=\int d^3\bm r\left\{\frac{|\nabla\psi|^2}{2m}+\alpha|\psi|^2\right\},\\
    \alpha=\frac{1}{mL}\left\{\frac{1}{g}-\int K^{\rm B}(\bm r)d^3\bm r\right\}.
\eea
From the bulk free energy above, the bulk critical temperature $T_{c0}$ and the bulk coherence length $\xi_{00}$ at zero temperature and zero field are determined by
\be
g\left.\int K^{\rm B}(\bm s)d^3\bm s\right|_{T=T_{c0}}=1,\quad \xi_{00}=0.18\frac{\hbar v_F}{k_{\rm B}T_{c0}},
\ee
where $v_F$ is the bulk Fermi velocity.

We consider the semi-infinite superconductor $\Sigma=\{(x,y,z)|z\leq 0\}$, and the surface $\partial\Sigma=\{(x,y,0)\}$ contains surface states $\phi^{\rm S}$ whose size of extension along $z$-axis is much smaller than $\xi_{00}$. As a result, the translation symmetries along in-plane $x,y$ directions are preserved, while the translation symmetry along $z$-axis is broken by the planar defect of the surface $z=0$.
In this case, along $z$-axis we need to perform the Laplace transform
\bea
    \hat{G}(\bm\rho,p)=\int_{-\infty}^{0}G(\bm\rho,z)e^{pz}dz
\eea
In the following we suppress $\bm\rho$ and write $\hat{G}(\bm\rho,p)=\hat{G}(p), G(\bm\rho,z)=G(z)$ for short.
In the GL regime, we focus on the long-range physics, which is $p\to 0$ in the Laplace domain $\hat{G}(p)\approx\hat{G}(0)$. Then we perform the inverse Laplace transform and get
\bea
    G(z)=\mathcal{L}^{-1}\{\hat{G}(p)\}=\hat{G}(0)\mathcal{L}^{-1}\{1\}=\hat{G}(0)\delta(z).
\eea
Thus the surface free energy reads
\bea
    F_s=\int d^3\bm r \Delta^*(\bm r)G(\bm r)=\int_{z=0} d^2\bm\rho\Delta^{*}(\bm\rho)\hat{G}(0),
\eea
where the zeroth order Laplace transform is just the integral over the lower $z$-axis
\bea
    \hat{G}(0)&=&\int_{-\infty}^{0}G(\bm\rho,z)dz\\\nonumber
    &=& \int_{-\infty}^{0}dz\int d^3\bm s[K^{\rm B}(\bm\rho-\bm s)-K(\bm\rho,\bm s)]\Delta(\bm s).
\eea

In terms of density of states $N,N_0$, bulk critical temperature $T_{c0}$, and Debye frequency $\omega_D$, we have
\bea
    \int {d^3\bm s}K(\bm r,\bm s)&=&N(\bm r)\log\frac{1.14\hbar\omega_D}{k_{\rm B}T_{c0}},\\\nonumber
    \int {d^3\bm s}K^{\rm B}(\bm r-\bm s)&=&N_0\log\frac{1.14\hbar\omega_D}{k_{\rm B}T_{c0}}=\frac{1}{g}.
\eea
Here we define the norm squared $|\phi|^2\equiv{\phi^{\dagger}\phi}$, then $N(\bm r)=\sum_a|\phi_{a}(\bm r)|^2\delta(\xi_a)$ is the local density of states, and $N_0=\underset{z \to -\infty}{\lim}N(\bm r)$ is the bulk density of states.
Finally we arrive at the expression of the surface free energy in the GL regime
\bea
    F_s&=&\int d^2\bm r U(\bm r)|\psi(\bm r)|^2,\\ \nonumber
    U(\bm r)&=&\int_{-\infty}^{0}\frac{\Delta(\bm r,z)}{\Delta_0}\left\{1-\frac{N(\bm r,z)}{N_0}\right\}\frac{dz}{mgL}.
\eea
where $\Delta_0=\Delta(\bm r)|_{z=0}$ is the pairing potential on the surface. In the clean limit, $N(\bm r,z)$ is a constant in the $xy$-plane on the scale of bulk coherence length, and $U(\bm r)=\sigma$ is spatially constant. 

By minimal coupling principle, we replace $\nabla$ by covariant derivative $\bm D=\nabla-2ie\bm A$ and arrive at the GL free energy
\bea
    F&=&F_0+F_s,\\\nonumber
    F_0&=&\int_{V} d^3\bm r\left\{\frac{|\bm D\psi|^2}{2m}+\alpha|\psi|^2+\frac{\beta}{2}|\psi|^4+\frac{1}{2}(\bm B-\bm H)^2\right\},\\\nonumber
    F_s&=&{\sigma}\int_{\partial V} |\psi|^2 d^2\bm r.
\eea

\subsection{Order parameter spatial profiles}
By minimizing the GL free energy with respect to $\psi$, we derive the GL equation
\bea
\left(-\frac{\bm D^2}{2m}+\alpha +\beta|\psi|^2\right)\psi =0,\ (\bm r\in V)
\eea
and the de Gennes boundary condition
\bea
    \hat{\bm n}\cdot\bm D\psi =\frac{\psi}{l_s},\ (\bm r\in\partial V)
\eea
where $l_{s}=1/(2m|\sigma|)>0$ is the extrapolation length denoting the thickness of surface superconductivity.

Now we solve the nonlinear GL equation under the generalized de Gennes boundary condition in the absence of external magnetic fields $\bm A=\bm 0$.
We introduce the bulk order parameter $\psi_0\equiv\sqrt{|\alpha|/\beta}$, the bulk coherence length $\xi_{0}\equiv\sqrt{1/(2m|\alpha|)}$, and the dimensionless temperature $t=(T-T_{c0})/(T_c-T_{c0})$, then the integral to the above equation reads
\be
-\xi_{0}^2\psi_0^2\left(\frac{d\psi}{dz}\right)^2+(\psi^2+\psi_0^2 {\rm sgn}\alpha )^2=C,
\ee
where the integral constant $C$ depends on the temperature $T$.

In the regime $T_{c0}<T<T_c$, most of the superconductor could not be superconducting except the surface layer. Deep in the bulk ($z\to -\infty$), $\psi\to 0$, $\alpha>0$ and hence $C=\psi_0^4>0$. The order parameter is solved as
\be\label{eq_op}
\psi(z) =\sqrt{2}\psi_0{{\rm csch}\left(\frac{|z|}{\xi_{0}} +\theta\right)},\quad (T_{c0}<T<T_c)
\ee
where $\theta ={\rm arctanh}\sqrt{|t|}$.
At temperatures $T\lesssim T_{c}$ and/or in the long range $|z|\gg\xi$, superconductivity decays exponentially $\psi(z)\sim\exp\left(-{|z|}/{\xi_{0}}\right)$ with bulk coherence length as the decay length.
In particular when $T=T_c$, $\xi_{0}=l_s$, superconductivity is localized $\psi\propto\exp(z/{l_s})$ in the surface layer with thickness $\sim l_s$.
At $T=T_{c0}$, the surface superconductivity reduces to power-law-like decay into the bulk $\psi(z)\propto({{|z|}/{l_s}+1})^{-1}.$  

In the regime $T<T_{c0}$, however, the bulk is already superconducting, and $\psi$ is mostly uniform except the surface layer. Deep in the bulk ($z\to -\infty$), $\psi\to\psi_0$, $\alpha<0$ and hence $C=0$. The order parameter is
\be
\psi(z) ={\psi_0}{\coth\left(\frac{|z|}{\sqrt{2}\xi_{0}}+\varphi\right)},\quad (T<T_{c0})
\ee
where $\varphi =\frac{1}{2}{\rm arcsinh} \sqrt{2|t|}$.
Below $T_{c0}$ and in the long range $|z|\gg\xi$, superconductivity is asymptotically uniform $\psi(z)\approx\psi_0$.

When $T\to T_c$, $t\to 1$ and Eq. (\ref{eq_op}) becomes Eq. (\ref{eq_phi})
\be
\psi(z)=\phi\exp\left(-\frac{|z|}{l_s}\right),\quad\phi=\sqrt{\frac{2\alpha_0(T_c-T)}{\beta}}.
\ee

\subsection{Meissner effect of field}
By minimizing the GL free energy with respect to $\bm A$, the Ampere's law is derived
\bea
\nabla\times\bm B=\bm j,\ (\bm r\in V)
\eea
with the Ampere's continuity boundary condition
$\hat{\bm n}\times\left(\bm H-\bm B\right)=\bm 0,\ (\bm r\in\partial V)$
and the supercurrent density
\be
\bm j=\frac{2e}{m} {\rm Im}(\psi^{*}\bm D\psi).
\ee

Now we solve the Ampere's law under the Ampere's continuity boundary condition in the presence of an external in-plane magnetic field $\bm H$.
We choose the gauge $\bm A=\bm H\times\hat{\bm n}f(z)$, then we have the Ampere's law
\be
f''(z)=\frac{2(1-t)}{\kappa^2l_s^2}\exp\left(\frac{2z}{l_s}\right)f(z)
\ee
with boundary condition $f'(0)=1$. The solution is found
\be
f(z)=A_0^{-1}{I_0\left[2\delta\exp\left(\frac{z}{l_s}\right)\right]},\ A_0=\delta I_{1}(2\delta),
\ee
where $\delta\equiv{\sqrt{2(1-t)}}/{\kappa}$ is a dimensionless quantity depending on temperature.

The internal magnetic field is
\bea
\bm B&=&\nabla\times\bm A=\bm Hf'(z),\\
f'(z)&=&\frac{\delta}{A_0l_s}\exp\left(\frac{z}{l_s}\right){I_1\left[2\delta\exp\left(\frac{z}{l_s}\right)\right]}.
\eea

Notice that $\delta\equiv{\sqrt{2(1-t)}}/{\kappa}$ is small near $T_c$ in type-II superconductors, we have
\be
A_0f(z)=1+\delta^2\exp\left(\frac{2z}{l_s}\right)+O(\delta^4).
\ee
Hence we find the internal magnetic field reads
\be
\bm B=\bm H\left\{e^{2z/l_s}+\frac{\delta^2}{2}e^{2z/l_s}\left(e^{2z/l_s}-1\right)\right\}+O(\delta^4).
\ee

\subsection{Surface supercurrent diode effect}
We choose the ansatz $\psi=\Delta\exp(z/l_s+i\bm q\cdot\bm r)$ with Cooper pair momentum $\bm q=(q_x,q_y)$ to account for finite supercurrent.
To determine the parameter $\Delta$, we need to take into account the nonlinear part of the GL equation, or equivalently we calculate the total free energy up to $|\Delta|^4$.
By minimizing the total free energy $F=F_0+F_s$, we find $|\Delta|^2=-2\tilde{\alpha}_{\bm q}/\beta$, where $\tilde{\alpha}_{\bm q}=\alpha_0(T-T_c)+|\bm q|^2/(2m)$.
Thus the supercurrent line density becomes
\bea
\bm J=\int_{-\infty}^{0}\bm jdz=I_0\left(\bm q-\bm q_0\right)(1-\xi^2 q^2),
\eea
where $\bm q_0=-\bm J_0/I_0=\bm H\times\hat{\bm n}/I_0$.
The dimensionless supercurrent diode coefficient is a function of the dimensionless quantity $x\equiv \xi \bm q_0\cdot\hat{\bm i}$
\be
\eta=
\begin{cases}
{x(9-x^2)}{(3+x^2)^{-3/2}} & |x|\leq 1\\
{\rm sgn}(x) & |x|>1
\end{cases}.
\ee
At weak fields $|x|\ll 1$, $\eta\approx\sqrt{3}x$ and we recover Eq. (\ref{eq_etas}).

\begin{figure}[h]
\includegraphics[width=\columnwidth]{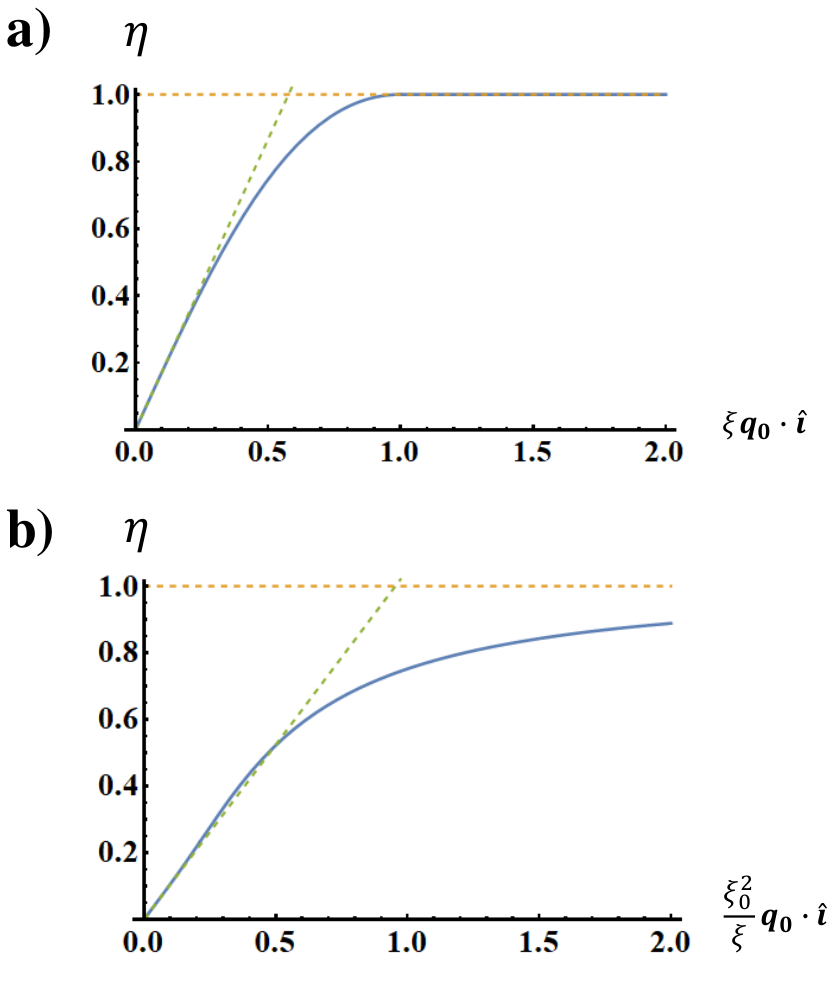}
\caption{Supercurrent diode coefficient $\eta$ as functions of the dimensionless quantity. a) In the surface supercurrent diode effect (SSDE), the dimensionless quantity is $\xi\bm q_0\cdot\hat{\bm i}$. At weak fields $|x|\ll 1$, $\eta\approx\sqrt{3}x$ as shown by the green dashed line. b) In the conventional supercurrent diode effect (CSDE), the dimensionless quantity is $\xi_{0}^2/\xi\bm q_0\cdot\hat{\bm i}$. Green dashed lines are linear approximation and orange dashed lines denote the perfect diode case $\eta=1$.}\label{figS}
\end{figure}

\section{Conventional supercurrent diode effect}
\subsection{Ginzburg-Landau coefficients}
We compute the the function $\sigma$ microscopically within mean field theory. The free energy is $F=\int f(\bm q) d^2\bm q $, and the free energy density can be expressed by the BdG Hamiltonian
$f=|\Delta|^2/V-T\int d^2\bm k{\rm tr}\log[1+e^{-\mathcal{H}/T}]$, where $V$ is the attractive interaction strength and tr denotes trace in spin space. 

By expansion in terms of $\Delta_{\bm q}$ we have
\begin{widetext}
\begin{eqnarray}\label{eq_f1c}
\sigma=N_0\log\frac{T}{T_c}+\sum_{\lambda=\pm}\oint_{{\rm FS}_{\lambda}}\frac{dk}{|\bm v^\lambda|} \left\{\phi \left(\frac{Q+\lambda\mathcal{E}_{+}}{2\pi T}\right)\cos^2\frac{\theta}{2}
+ \phi \left(\frac{Q+\lambda\mathcal{E}_{-}}{2\pi T}\right)\sin^2\frac{\theta}{2}\right\},
\end{eqnarray}
\end{widetext}
where $\lambda=\pm$ denote contributions from inner ($\lambda=+$) and outer ($\lambda=-$) Fermi surfaces respectively,
$\phi(x)={\rm Re}\left[\psi\left(\frac{1+ix}{2}\right)\right]-\psi\left(\frac{1}{2}\right),$
and $\psi$ is the digamma function. 

Here, $\bm v^\lambda=\partial_{\bm k}\xi_{\bm k}^{\lambda}$ is the electron velocity,
$Q=\bm v^{\lambda}\cdot\bm q$ is the depairing energy of finite momentum pairing, $ \mathcal{E}_{\pm}=|\bm h_{+}|\pm |\bm h_{-}| $ is the depairing energy of Zeeman splitting for inter- ($+$) or intra-pocket ($-$) Cooper pairs with $\bm h_{\pm}=\bm B+\bm g_{\frac{1}{2}\bm q\pm\bm k}$, and the angle $ \theta=\langle\bm h_{+},\bm h_{-}\rangle $ between $\bm h_{\pm}$ controls the ratio between inter- or intra-pocket Cooper pairs.
Supercurrent affects $\bm q$ and hence depairing energy $Q$, while magnetic field $\bm B$ together with SOC affects depairing energies $\mathcal{E}_{\pm}$ and angle $\theta$. 

Near $T_c$, the temperature dependence of $\sigma_{\bm q}$ can be captured by the first term $N_0\log(T/T_c)$, and we can set $T=T_c$ in the Fermi surface integrals. To evaluate the integral, notice that the field is weak $H\ll H_P$, one can expand the special function 
\begin{eqnarray}
\phi(x)=2.10x^2-2.01x^4+2.00x^6+O(x^8)
\end{eqnarray}
and then integrate order by order to obtain GL coefficients.

\subsection{Nonreciprocal critical current and polarity-dependent critical field}
When the expansion of $\sigma$ near its minimum $\bm q_0$ reads 
\begin{eqnarray}
\sigma=\sigma_0+a\delta q_{\parallel}^2-b\delta q_{\parallel}^3,
\end{eqnarray}
where $a>0$ and $\delta q_{\parallel}=(\bm q-\bm q_0)\cdot\hat{\bm q}_0$.
We find the supercurrent is
\begin{eqnarray}\nonumber
\gamma J_{\parallel}/e=|\sigma_{\bm q}|\partial_{\parallel}\sigma_{\bm q}=2 a \sigma_0 \delta q_{\parallel} -3 \left(\sigma_0 b\right)\delta q_{\parallel}^2\\
+2 a^2 \delta q_{\parallel}^3 - 5 (a b) \delta q_{\parallel}^4 +3b^2\delta q_{\parallel}^5.
\end{eqnarray}

Notice that to the leading order of $|\sigma_0|^{\frac{1}{2}}$, critical currents $\pm J_{c}^{\pm}$ correspond to $\delta q_{\parallel}=\mp \delta q_c$ respectively, where $\delta q_c=\sqrt{|\sigma_0|/3a}$. Then we have
\be\label{eq_jc}
\gamma J_{c}^{\pm}/e=\frac{4 |\sigma_0|^{3/2}}{9 a}\left(\sqrt{3a^3}\pm b\sqrt{|\sigma_0|}\right) +O(|\sigma_0|^{5/2}).
\ee

To include higher order contributions, the supercurrent diode coefficient is
\begin{equation}
    \eta=g\left(x\right),\quad x=\sqrt{\frac{|\sigma_0|}{a^3}} {b}.
\end{equation}
The special function is
\be
g(x)=\frac{J(x)-J(-x)}{J(x)+J(-x)},
\ee
\be
J(x)=[Q(x)-\frac{3}{2}xQ(x)][-1+Q(x)^2-xQ(x)^3]
\ee
\be
Q(x)=\frac{1}{3 x}-\frac{\sqrt{n}}{2 x}-\frac{1}{2} \sqrt{{\frac{4}{5\sqrt{n}}}\left(1-\frac{2}{27 x^2}\right)-\frac{n}{x^2}+\frac{8}{15 x^2}}
\ee
\be
n=\frac{2}{15}\left(z+1/{z}+\frac{4}{3}\right),
\ee
\be
z=\sqrt[3]{t+\sqrt{t^2-1}},\quad
t=\frac{135}{4}x^4-5 x^2+1. 
\ee
When $|x|<\frac{2}{3\sqrt{3}}$, $\frac{22}{27}<t<1$ and $z$ is complex. Denote $t=\cos\theta$, then $z=e^{i\theta/3}$ and $n=\frac{4}{15}(\cos\frac{\theta}{3}+\frac{2}{3})$ is real.
Since $g(x)\approx x/\sqrt{3}$ we get the leading order contribution $\eta=\sqrt{\frac{|\sigma_0|}{3a^3}} {b}$.

{The expansion of $\sigma_{\bm q}$ near its minimum $\bm q_0$ in general can be anisotropic
\begin{eqnarray}\nonumber
\sigma_{\bm q+\bm q_0}=\sigma_0+a(1+\epsilon) q_x^2+a(1-\epsilon)q_y^2 +2a\eta q_x q_y \\\nonumber
-(b_1 q_x^3 +b_2 q_y^3 +b_3 q_x q_y^2+b_4 q_x^2 q_y),
\end{eqnarray}
where $a>0$ and $\epsilon^2+\eta^2<1$ for stability. The supercurrent diode coefficient for supercurrent $\bm J=J(\cos\theta,\sin\theta)$ can be worked out as
\begin{widetext}
\begin{eqnarray}\label{eq_diode}
\eta =\sqrt{\frac{|\sigma_0|}{3a^3}}\frac{\left(\frac{b_1+b_3}{2}+\frac{b_1-b_3}{2}\cos2\theta\right)\cos\theta+\left(\frac{b_2+b_4}{2}-\frac{b_2-b_4}{2}\cos2\theta\right)\sin\theta}{(1+\epsilon\cos 2\theta+\eta\sin 2\theta)^{3/2}}.
\end{eqnarray}
\end{widetext}}


\begin{thebibliography}{100}

\bibitem{Waka} R. Wakatsuki, Y. Saito, S. Hoshino, Y. M. Itahashi, T. Ideue, M. Ezawa, Y. Iwasa, and N. Nagaosa, \textit{Nonreciprocal charge transport in noncentrosymmetric
superconductors}, Sci. Adv. \textbf{3}, e1602390 (2017).

\bibitem{Qin} F. Qin, W. Shi, T. Ideue, M. Yoshida, A. Zak, R. Tenne, T. Kikitsu, D. Inoue, D. Hashizume and Y. Iwasa, \textit{Superconductivity in a chiral nanotube}, Nat. Commun. \textbf{8}, 14465 (2017).

\bibitem{Yasu} Kenji Yasuda, Hironori Yasuda, Tian Liang, Ryutaro Yoshimi, Atsushi Tsukazaki, Kei S. Takahashi, Naoto Nagaosa, Masashi Kawasaki and Yoshinori Tokura, \textit{Nonreciprocal charge transport at topological insulator/superconductor
interface}, Nat. Commun. \textbf{10}, 2734 (2019).

\bibitem{Hoshi} S. Hoshino, R. Wakatsuki, K. Hamamoto, and N. Nagaosa, \textit{Nonreciprocal charge transport in two-dimensional noncentrosymmetric superconductors}, Phys. Rev. B \textbf{98}, 054510 (2018).

\bibitem{Ando} Fuyuki Ando, Yuta Miyasaka, Tian Li, Jun Ishizuka, Tomonori Arakawa, Yoichi Shiota, Takahiro Moriyama, Youichi Yanase and Teruo Ono, \textit{Observation of superconducting diode effect}, Nature \textbf{584}, 373 (2020).

\bibitem{Chris} Christian Baumgartner, Lorenz Fuchs, Andreas Costa, Simon Reinhardt, Sergei Gronin, Geoffrey C. Gardner, Tyler Lindemann, Michael J. Manfra, Paulo E. Faria Junior, Denis Kochan, Jaroslav Fabian, Nicola Paradiso and Christoph Strunk, \textit{Supercurrent rectification and magnetochiral effects in symmetric Josephson junctions}, Nature Nanotechnology \textbf{17}, 39 (2022).

\bibitem{Banabir} Banabir Pal, Anirban Chakraborty, Pranava K. Sivakumar, Margarita Davydova, Ajesh K. Gopi, Avanindra K. Pandeya, Jonas A. Krieger, Yang Zhang, Mihir Date, Sailong Ju, Noah Yuan, Niels B. M. Schröter, Liang Fu and Stuart S. P. Parkin, \textit{Josephson diode effect from Cooper pair momentum in a topological semimetal}, Nature Physics \textbf{18}, 1228 (2022).

\bibitem{Lorenz} Lorenz Bauriedl, Christian Bäuml, Lorenz Fuchs, Christian Baumgartner, Nicolas Paulik, Jonas M. Bauer, Kai-Qiang Lin, John M. Lupton, Takashi Taniguchi, Kenji Watanabe, Christoph Strunk and Nicola Paradiso, \textit{Supercurrent diode effect and magnetochiral anisotropy in few-layer NbSe$_2$}, Nat. Comm. \textbf{13}, 4266 (2022).

\bibitem{Diez} J. Diez-Merida, A. Diez-Carlon, S. Y. Yang, Y.-M. Xie, X.-J. Gao, K. Watanabe, T. Taniguchi, X. Lu, K. T. Law, and Dmitri K. Efetov, \textit{Magnetic Josephson Junctions and Superconducting Diodes in Magic Angle Twisted Bilayer Graphene}, arXiv:2110.01067.

\bibitem{LinJ} Jiang-Xiazi Lin, Phum Siriviboon, Harley D. Scammell, Song Liu, Daniel Rhodes, K. Watanabe, T. Taniguchi, James Hone, Mathias S. Scheurer and J.I.A. Li, \textit{Zero-field superconducting diode effect in small-twist-angle trilayer graphene}, Nat. Phys. \textbf{18}, 1221 (2022). 

\bibitem{EdelEE} Victor M. Edelstein, \textit{Characteristics of the Cooper pairing in two-dimensional noncentrosymmetric electron systems}, Sov. Phys. JETP \textbf{68} (6), (1989).

\bibitem{EdelEE1} Victor M. Edelstein, \textit{Spin polarization of conduction electrons
induced by electric current in two-dimensional asymmetric electron systems}, Solid State Commun. \textbf{73}, 233 (1990).

\bibitem{EdelME} Victor M. Edelstein, \textit{Magnetoelectric Effect in Polar Superconductors}, Phys. Rev. Lett. \textbf{75}, 2004 (1995).

\bibitem{EdelSDE} Victor M. Edelstein, \textit{The Ginzburg-Landau equation for superconductors of polar symmetry}, J. Phys.: Condens. Matter 8, \textbf{339} (1996).

\bibitem{Yuan} Noah F. Q. Yuan and Liang Fu, \textit{Supercurrent diode effect and finite-momentum superconductors}, PNAS \textbf{119} (15) e2119548119 (2022).

\bibitem{Akito} Akito Daido, Yuhei Ikeda, and Youichi Yanase, \textit{Intrinsic Superconducting Diode Effect}, Phys. Rev. Lett. \textbf{128}, 037001 (2022).

\bibitem{James} James Jun He, Yukio Tanaka, and Naoto Nagaosa, \textit{A phenomenological theory of superconductor diodes}, New J. Phys. \textbf{24}, 053014 (2022).

\bibitem{Harley} Harley D Scammell, J I A Li and Mathias S Scheurer, \textit{Theory of zero-field superconducting diode effect in twisted trilayer graphene}, 2D Mater. \textbf{9}, 025027 (2022).

\bibitem{Zhai} B. Zhai, B. Li, Y. Wen, F. Wu, and J. He, \textit{Prediction of ferroelectric superconductors with reversible superconducting diode effect}, Physical Review B \textbf{106}, L140505 (2022).

\bibitem{Ilic} S. Ili\'c and F. S. Bergeret, \textit{Theory of the Supercurrent Diode Effect in Rashba Superconductors with Arbitrary Disorder}, Phys. Rev. Lett. \textbf{128}, 177001 (2022).

\bibitem{Dimi1} V. Dimitrova, M. V. Feigel’man, \textit{Phase diagram of a surface superconductor in parallel magnetic field}, JETP Lett. \textbf{78}, 637 (2003).

\bibitem{Dimi2} V. Dimitrova, M. V. Feigel’man, \textit{Theory of a two-dimensional superconductor with broken inversion symmetry}, Phys. Rev. B \textbf{76}, 014522 (2007).

\bibitem{Samokhin} K. V. Samokhin, \textit{Upper critical field in noncentrosymmetric superconductors}, Phys. Rev. B \textbf{78}, 224520 (2008).

\bibitem{TI} C.-X. Liu, X.-L. Qi, H. Zhang, X. Dai, Z. Fang, and Zhang S.-C. \textit{Model Hamiltonian for topological insulators}, {Phys. Rev. B} \textbf{82}, 045122 (2010).

\bibitem{Simonin} J. Simonin, \textit{Surface term in the superconductive Ginzburg-Landau free energy: Application to thin films}, {Phys. Rev. B} \textbf{33}, 7830(R) (1986).

\bibitem{Andryushin}  E. A. Andryushin, V. L. Ginzburg, and A. P. Silin, \textit{Boundary conditions in the macroscopic theory of superconductivity}, {Phys.-Usp.} \textbf{36}, 854 (1993).

\bibitem{deGennes} P. G. de Gennes, \textit{Superconductivity of metals and alloys} (W. A. Benjamin, New York, 1966).

\bibitem{saint} D. Saint-James, and P. G. de Gennes, \textit{Onset of superconductivity in decreasing fields}, {Phys. Lett.} \textbf{7}, 306 (1963).

\bibitem{Noah} Noah F. Q. Yuan and X.-J. Chen, \textit{Critical field measure for topological}, arXiv:2209.00816.

\bibitem{Diehl} H. W. Diehl, \textit{The theory of boundary critical phenomena}, Int. J. Mod. Phys. B11: 3503 (1997).

\bibitem{Abrikosov1} A. A. Abrikosov, \textit{On the magnetic properties of superconductors of the second group}, Sov. Phys. JETP \textbf{5}, 1174 (1957).

\bibitem{Gang} Gang Xu, Biao Lian, Peizhe Tang, Xiao-Liang Qi, and Shou-Cheng Zhang, \textit{Topological superconductivity on the surface of Fe-based superconductors}, Phys. Rev. Lett. \textbf{117}, 047001 (2016).

\bibitem{Zhang} Peng Zhang, Koichiro Yaji, Takahiro Hashimoto, Yuichi Ota, Takeshi Kondo, Kozo Okazaki, Zhijun Wang, Jinsheng Wen, G. D. Gu, Hong Ding, and Shik Shin, \textit{Observation of topological superconductivity on the surface of iron-based superconductor}, Science \textbf{360}, 182 (2018).

\bibitem{Wang} Dongfei Wang, Lingyuan Kong, Peng Fan, Hui Chen, Shiyu Zhu, Wenyao Liu, Lu Cao, Yujie Sun, Shixuan Du, John Schneeloch, Ruidan Zhong, Genda Gu, Liang Fu, Hong Ding, and Hong-Jun Gao, \textit{Evidence for Majorana bound states in an iron-based superconductor}, Science \textbf{362}, 333 (2018).

\bibitem{Kong} Lingyuan Kong, Shiyu Zhu, Michał Papaj, Lu Cao, Hiroki Isobe, Wenyao Liu, Dongfei Wang, Peng Fan, Hui Chen, Yujie Sun, Shixuan Du, John Schneeloch, Ruidan Zhong, Genda Gu, Liang Fu, Hong-Jun Gao, and Hong Ding, \textit{Half-integer level shift of vortex bound states in an iron-based superconductor}, Nat. Phys. \textbf{15}, 1181 (2019).





\bibitem{Gao} Miao Gao, Fengjie Ma, Zhong-Yi Lu, and Tao Xiang, \textit{Surface structures of ternary iron arsenides AFe$_2$As$_2$ (A=Ba, Sr, or Ca)}, Phys. Rev. B \textbf{81}, 193409 (2010).

\bibitem{Heumen} Erik van Heumen, Johannes Vuorinen, Klaus Koepernik, Freek Massee, Yingkai Huang, Ming Shi, Jesse Klei, Jeroen Goedkoop, Matti Lindroos, Jeroen van den Brink, and Mark S. Golden, \textit{Existence, character and origin of surface-related bands in the high temperature iron pnictide superconductor BaFe$_{2-x}$Co$_{x}$As$_{2}$}, Phys. Rev. Lett. \textbf{106}, 027002 (2011).

\bibitem{Par} Christopher T. Parzyck , Brendan D. Faeth , Gordon N. Tam, Gregory R. Stewart, and Kyle M. Shen, \textit{Enhanced surface superconductivity in
Ba(Fe$_{0.95}$Co$_{0.05}$)$_2$As$_2$}, Appl. Phys. Lett. \textbf{116}, 062601 (2020).

\bibitem{Matsu} M. Matsumoto, C. Belardinelli, and M. Sigrist, \textit{Upper critical field of the 3 Kelvin phase in Sr$_2$RuO$_4$}, J. Phys. Soc. Jpn. \textbf{72}, 1623 (2003).

\bibitem{Khl} I. N. Khlyustikov, and A. I. Buzdin, \textit{Twinning-plane superconductivity}, Adv. Phys. \textbf{36}, 271 (1987).

\bibitem{Abrikosov2}  A. A. Abrikosov, \textit{Concerning surface superconductivity in strong magnetic fields}, Sov. Phys. JETP \textbf{20}, 480 (1965).



\bibitem{Yuan1} Noah F. Q. Yuan and Liang Fu, \textit{Topological metals and finite-momentum superconductors}, PNAS \textbf{118} (3), e2019063118 (2021).

\end{thebibliography}
\end{document}